\documentclass[aip,jap,preprint,amsmath,amssymb]{revtex4-1}
\usepackage{graphicx}% Include figure files
\usepackage{dcolumn}% Align table columns on decimal point
\usepackage{bm}% bold math

\newcommand{\BTO}{$\text{BaTiO}_3$}
\newcommand{\STO}{$\text{SrTiO}_3$}
\newcommand{\PTO}{$\text{PbTiO}_3$}

\begin{document}

\title{Positive effective $Q_{12}$ electrostrictive coefficient in perovskites}

\author{Alexander Kvasov}
\author{Alexander K. Tagantsev}

\affiliation{%
Ceramics Laboratory\\
\'Ecole Polytechnique F\'ed\'erale de Lausanne \\
CH-1015 Lausanne, Switzerland
}%
\date{\today}
\pacs{77.55.-g, 77.65.-j, 77.80.Jk}
%77.55.-g	Dielectric thin films
%77.65.-j	Piezoelectricity and electromechanical effects
%77.80.Jk	Relaxor ferroelectrics

\begin{abstract}
It is demonstrated that for classical perovskites such as \BTO{}, \STO{} and \PTO{} electrostrictive strain induced by an electric field may not obey traditionally considered "extension along the field,
contraction perpendicular to it" behavior if a sample is cut obliquely to the cubic crystallographic directions.
\end{abstract}

\maketitle

\section{Introduction}\label{introduction}
% Q is important, can be used in resonators / TFBARs
To exhibit a deformation under the application of an electric field, is a common property of solids.
Historically, this phenomena was called electrostriction and it implied both a linear converse piezoelectric effect and a quadric effect with respect to polarization, i.e. electrostriction.
The electrostrictive effect, being quadratic, is typically weaker than the piezoelectric one.
At the same time, in materials of interest for applications, primarily ferroelectrics, the piezoelectric effect has its disadvantages, for example it shows hysteresis which is undesirable for  actuator applications \cite{Uchino_book}, while electrostriction in centrosymmetric materials is free of this drawback.
Electrostriction plays an essential role in physics and applications of solids.
First of all, electrostriction is of importance for thin film strain engineering \cite{Schlom_2007}.
In relaxor ferroelectrics, in view of their high dielectric constant, the electrostrictive effect can be used for actuator applications \cite{Uchino_book}.
Electrostriction is important for tunable thin film bulk acoustic resonators (TFBARs) based on ferroelectric films where it is responsible for the dc-field-induced piezoelectricity in the material \cite{Noeth_jap_2008, Berge_jap_2008}.

It is generally believed that in perovskite ferroelectrics the electrostrictive effect leads to expansion of the sample along the applied electric field and its contraction in the perpendicular direction.
However, this is not a general rule for ferroelectrics.
For example, some of the ferroelectric polymers from the polyvinylidene fluoride (PVDF) family can exhibit the opposite behavior \cite{Fukada, Kochervinskii}.
Here a reasonable question arises; is the "extension along the field, contraction perpendicular to it" rule general for perovskites or is it rather a feature of the crystallographic orientation of the sample?
This paper addresses this question to show that the "extension along the field, contraction perpendicular to it" rule can be readily violated if the sample is cut obliquely to the cubic crystallographic directions.

\section{Positive effective $Q_{12}$ electrostriction}\label{main}

\begin{figure}[!ht]
\includegraphics[width=0.48\textwidth]{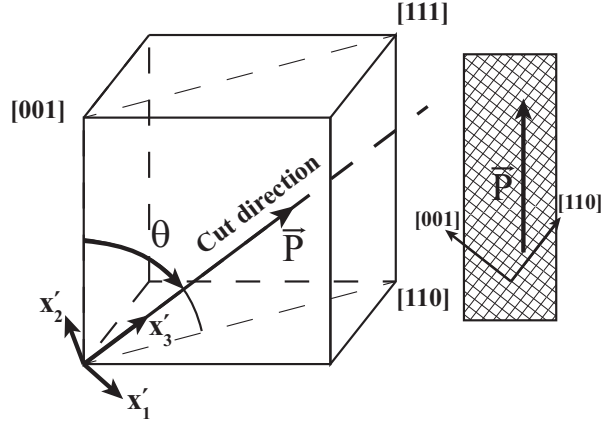}
\caption{Direction of polarization described by angle $\theta$ and the rotated reference frame
(${x'}_1,{x'}_2,{x'}_3$) associated with it. The polarization direction changes in plane going from [001]
to [111] and then to the [110] direction.}
\label{fig_001-111}
\end{figure}

To answer this question, let us consider a bar of perovskite material in the cubic paraelectric phase cut
obliquely in a direction which differs by angle $\theta$ from the [001] crystallographic axis (see Fig.
\ref{fig_001-111}). The oblique cut direction changes in the (110)-plane and passes from [001] to [111] and then to [110]-directions. To explore the electrostrictive strain behavior, one applies an electrical field
$\vec{E}$ which in turn induces polarization $\vec{P}$ parallel to the field ( $\vec{P} = \chi \vec{E}$ for cubic material) in the oblique cut direction.
The polarization $\vec{P}$ in the sample
\begin{equation}\label{eq_P}
\vec{P} = \begin{pmatrix} \frac{\sqrt{2}}{2} \sin{\theta} \\ \frac{\sqrt{2}}{2} \sin{\theta} \\
\cos{\theta}  \end{pmatrix}
\end{equation}
in turn induces electrostrictive strain $\epsilon_{kl}$:
\begin{equation}\label{eq_Q_def}
\epsilon_{kl} = Q_{ijkl} P_i P_j
\end{equation}
where $Q_{ijkl}$ is the electrostrictive tensor. The Einstein dummy suffix summation convention is adopted.
We now change to the Voigt matrix notation
\begin{eqnarray}\label{eq_epsilon_voigt}
& \epsilon_{ij} &= \epsilon_{n} \text{ for } n=1,2,3 \nonumber \\
& \epsilon_{ij} &= \frac{\epsilon_{n}}{2} \text{ for } n=4,5,6, \nonumber
\end{eqnarray}
defining electrostrictive tensor $Q_{mn}$ according to the Landolt-Bornstein reference book \cite{Landolt-Bornstein}:
\begin{eqnarray}\label{eq_Q_voigt}
& Q_{ijkl} &= Q_{mn} \text{ for } n=1,2,3 \nonumber \\
& Q_{ijkl} &= \frac{Q_{mn}}{2} \text{ for } n=4,5,6. \nonumber
\end{eqnarray}
In Voigt notation the electrostrictive strain $\epsilon$ appeared in the sample reads as
\begin{equation}\label{eq_epsilon}
\epsilon =
\begin{pmatrix} \epsilon_1 \\ \epsilon_1  \\ \epsilon_3 \\ \epsilon_4 \\ \epsilon_4 \\ \epsilon_6
\end{pmatrix},
\end{equation}
where
\begin{eqnarray}\label{eq_epsilon_legend}
& \epsilon_1 &= P^2 \left(\frac{1}{2} (Q_{11} + Q_{12}) \sin^2{\theta} + Q_{12} \cos^2{\theta} )\right)
\nonumber \\
& \epsilon_3 &= P^2 \left( Q_{11} \cos^2{\theta} + Q_{12} \sin^2{\theta} \right) \nonumber \\
& \epsilon_4 &= \frac{\sqrt{2}}{2} P^2 Q_{44} \cos{\theta} \sin{\theta} \nonumber \\
& \epsilon_6 &= \frac{1}{2} P^2 Q_{44} \sin^2{\theta} \nonumber
\end{eqnarray}
In the Cartesian reference frame (${x'}_1,{x'}_2,{x'}_3$) with the ${x'}_3$ axis parallel to the direction of the oblique cut (see Fig. \ref{fig_001-111}), the strain tensor has the form:
\begin{equation}\label{eq_epsilon_rotated}
\epsilon' =
\begin{pmatrix} P^2 Q_{12}^{\theta(1)} \\ P^2 Q_{12}^{\theta(2)}  \\ P^2 Q_{11}^\theta \\ P^2 Q_{44}^\theta
\\ 0 \\ 0
\end{pmatrix}
\end{equation}
where $Q_{11}^\theta$, $Q_{12}^{\theta(1)}$, $Q_{12}^{\theta(2)}$, and $Q_{44}^\theta$ are the effective
electrostrictive components.
$\epsilon'_{3} = P^2 Q_{11}^\theta$ is parallel to the polarization and $\epsilon'_{1} = P^2
Q_{12}^{\theta(1)}$ and $\epsilon'_{2} = P^2 Q_{12}^{\theta(2)}$ are perpendicular to it.
For $\theta=0$, for typical cubic perovskites \BTO{}, \STO{}, and \PTO{} $Q_{11}^\theta=Q_{11}>0$ while $Q_{12}^{\theta(1)}= Q_{12}^{\theta(2)}= Q_{12}<0$ (see Table \ref{table_q}) and the "extension along the field, contraction perpendicular to it" rule holds.
In answering the question of whether this rule is violated for the oblique orientation of the sample, we are interested in whether the signs of $Q_{11}^\theta$, $Q_{12}^{\theta(1)}$, and $Q_{12}^{\theta(2)}$ depend on the value of angle $\theta$.

\begin{figure}[!ht]
\includegraphics[width=0.45\textwidth]{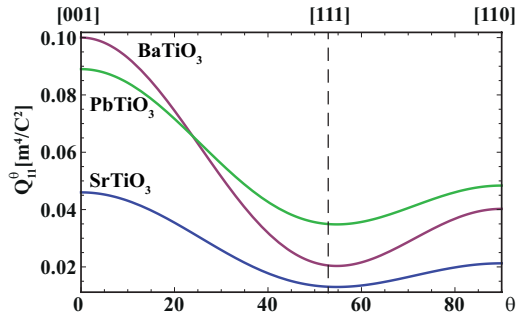}
\caption{Effective $Q_{11}^\theta$ coefficient as a function of the cut direction $\theta$ for
three perovskites materials \BTO{} (pink), \STO{} (blue) and \PTO{} (green). The data for electrostrictive constants are taken from Table \ref{table_q}.}
\label{fig_q11}
\end{figure}

\begin{figure}[!ht]
\includegraphics[width=0.45\textwidth]{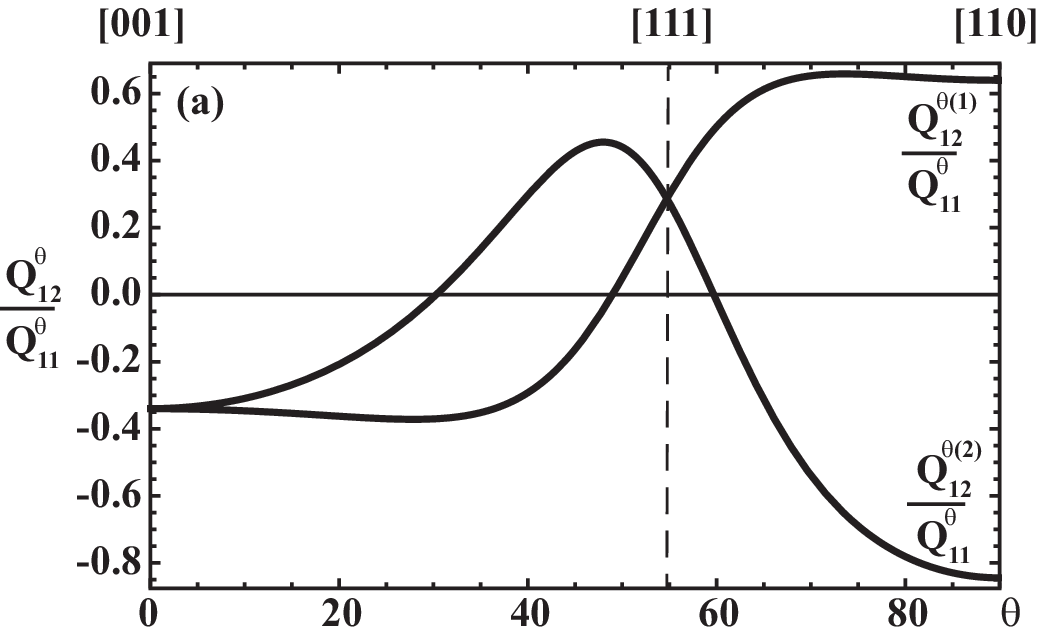}
\includegraphics[width=0.45\textwidth]{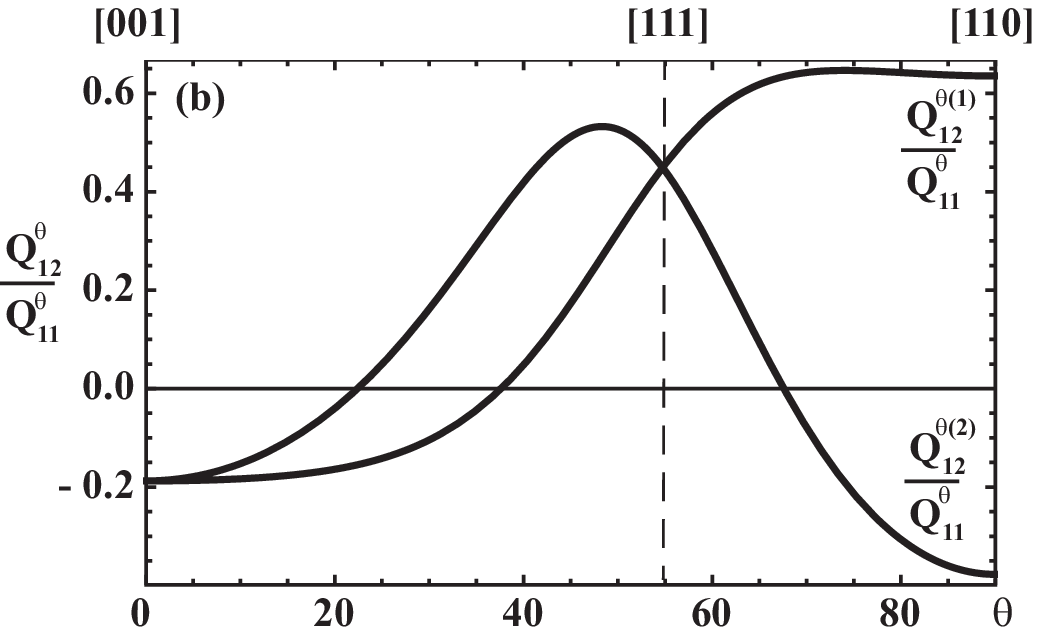}
\includegraphics[width=0.45\textwidth]{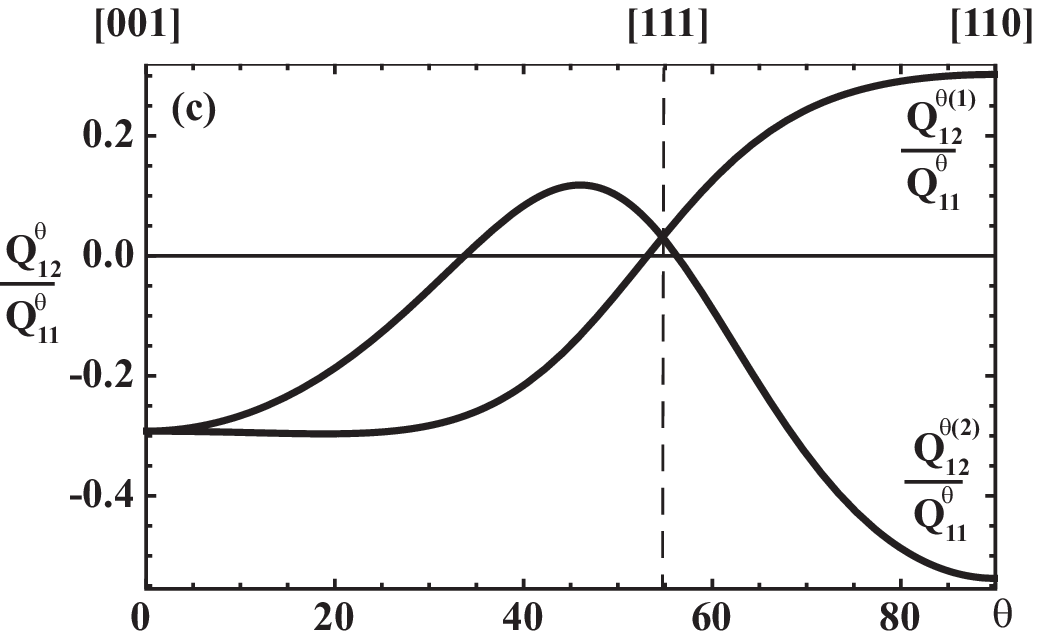}
\caption{Dependence of $\frac{Q_{12}^\theta}{Q_{11}^\theta}$ ratio on the cut direction $\theta$.
$Q_{11}^\theta$ - longitudinal effective electrostrictive coefficient, $Q_{12}^{\theta (1)}, Q_{12}^{\theta
(2)}$ - two transversal effective electrostrictive coefficients (see \eqref{eq_epsilon_rotated}).
(a) plotted for \BTO{} with coefficients from \cite{Yamada_BTO}.
(b) for \STO{} from \cite{Tagantsev_STO}.
(c) for \PTO{} from \cite{Pertsev_Ferroelectrics_1999}.
}
\label{fig_positive_q}
\end{figure}

We start with an issue relevant to the problem.
The field-induced relative change of the sample volume is given by the trace of the deformation tensor, which is independent of the direction of the applied field:
\begin{equation}\label{eq_trace}
Tr(\epsilon_{ij}) = P^2(Q_{11}^\theta + Q_{12}^{\theta(1)} + Q_{12}^{\theta(2)}) = P^2(Q_{11} + 2 Q_{12}) =
\text{const}.
\end{equation}
Thus, in principle, if one cuts the sample obliquely with respect to the cubic crystallographic directions $Q_{12}^{\theta}$ can become positive at the expense of a reduction in $Q_{11}^{\theta}$ while keeping the sum $Q_{11}^\theta + Q_{12}^{\theta(1)} + Q_{12}^{\theta(2)}$ constant and
breaking the classical "extension along the field, contraction perpendicular to it" behavior.
Indeed, for three perovskites materials \BTO{}, \STO{}, and \PTO{}, such a reduction in $Q_{11}^\theta$ is possible as clear from Fig. \ref{fig_q11}.
\begin{table}[!ht]
\begin{tabular}{|c|c|c|c|}
\hline
 & \BTO{} \cite{Yamada_BTO} & \STO{} \cite{Tagantsev_STO} & \PTO{} \cite{Pertsev_Ferroelectrics_1999} \\
\hline
$Q_{11},\left[\frac{\text{m}^4}{\text{C}^2}\right]$ & 0.10 & 0.046 & 0.089 \\
\hline
$Q_{12},\left[\frac{\text{m}^4}{\text{C}^2}\right]$ & $-0.034$ & $-0.013$ & $-0.026$ \\
\hline
$Q_{44},\left[\frac{\text{m}^4}{\text{C}^2}\right]$ & 0.015 & 0.010 & 0.034 \\
\hline
\end{tabular}
\caption{Electrostrictive coefficients which were used for calculations, for \BTO{} coefficients were taken
from \cite{Yamada_BTO}, for \STO{} from \cite{Tagantsev_STO}, and \PTO{} from
\cite{Pertsev_Ferroelectrics_1999}. The $Q_{44}$ values are written taking into account the factor of 2 as
defined in Landolt–Bornstein \cite{Landolt-Bornstein}.}
\label{table_q}
\end{table}
It is seen in Fig. \ref{fig_q11} that while $Q_{11}^\theta$ always stays positive, at the same time, it decreases for all the considered perovskite materials.
Such a decrease corresponds to appreciable variations in $Q_{12}^{\theta}$ for all these materials, as seen from Fig. \ref{fig_positive_q}, where the dependences of the $\frac{Q_{12}^\theta}{Q_{11}^\theta}$ ratio on the cut direction $\theta$ are plotted.
For the cubic ferroelectrics considered, a positive  $Q_{12}^{\theta}$ can be found for certain directions.
Cuts close to the [111]-direction are of special interest.
Here one finds expansion of the sample in all directions.
In addition, $Q_{11}^{\theta}$ has a minimum for the [111]-direction which in turn gives a maximum for $Q_{12}^{\theta(1)} + Q_{12}^{\theta(2)}$.
For cuts close to the [110]-direction, instead of the classical "extension along the field, contraction perpendicular to it" behavior one finds a very different behavior where, under the field, the sample  expands in two dimensions and contracts in one (along the [001]-axis).

\section{Conclusions}\label{conclusions}
It was shown that classical cubic perovskite ferroelectrics exhibit deviations from the "extension along the field, contraction perpendicular to it" behavior if the sample is cut obliquely along cubic crystallographic directions.
Particularly, one can find electrostrictive "expansion in all dimensions" like for
the [111] cut or a situation where the sample contracts only in one dimension and expands in two.
The crystal chemistry behind such behavior is not clear for the moment.
This phenomenon may be of interest for actuators based on the electrostrictive effect.

\bibliography{ref}

%merlin.mbs aipnum4-1.bst 2010-07-25 4.21a (PWD, AO, DPC) hacked
%Control: key (0)
%Control: author (8) initials jnrlst
%Control: editor formatted (1) identically to author
%Control: production of article title (-1) disabled
%Control: page (0) single
%Control: year (1) truncated
%Control: production of eprint (0) enabled
\begin{thebibliography}{10}%
\makeatletter
\providecommand \@ifxundefined [1]{%
 \@ifx{#1\undefined}
}%
\providecommand \@ifnum [1]{%
 \ifnum #1\expandafter \@firstoftwo
 \else \expandafter \@secondoftwo
 \fi
}%
\providecommand \@ifx [1]{%
 \ifx #1\expandafter \@firstoftwo
 \else \expandafter \@secondoftwo
 \fi
}%
\providecommand \natexlab [1]{#1}%
\providecommand \enquote  [1]{``#1''}%
\providecommand \bibnamefont  [1]{#1}%
\providecommand \bibfnamefont [1]{#1}%
\providecommand \citenamefont [1]{#1}%
\providecommand \href@noop [0]{\@secondoftwo}%
\providecommand \href [0]{\begingroup \@sanitize@url \@href}%
\providecommand \@href[1]{\@@startlink{#1}\@@href}%
\providecommand \@@href[1]{\endgroup#1\@@endlink}%
\providecommand \@sanitize@url [0]{\catcode `\\12\catcode `\$12\catcode
  `\&12\catcode `\#12\catcode `\^12\catcode `\_12\catcode `\%12\relax}%
\providecommand \@@startlink[1]{}%
\providecommand \@@endlink[0]{}%
\providecommand \url  [0]{\begingroup\@sanitize@url \@url }%
\providecommand \@url [1]{\endgroup\@href {#1}{\urlprefix }}%
\providecommand \urlprefix  [0]{URL }%
\providecommand \Eprint [0]{\href }%
\providecommand \doibase [0]{http://dx.doi.org/}%
\providecommand \selectlanguage [0]{\@gobble}%
\providecommand \bibinfo  [0]{\@secondoftwo}%
\providecommand \bibfield  [0]{\@secondoftwo}%
\providecommand \translation [1]{[#1]}%
\providecommand \BibitemOpen [0]{}%
\providecommand \bibitemStop [0]{}%
\providecommand \bibitemNoStop [0]{.\EOS\space}%
\providecommand \EOS [0]{\spacefactor3000\relax}%
\providecommand \BibitemShut  [1]{\csname bibitem#1\endcsname}%
\let\auto@bib@innerbib\@empty
%</preamble>
\bibitem [{\citenamefont {Uchino}(2000)}]{Uchino_book}%
  \BibitemOpen
  \bibfield  {author} {\bibinfo {author} {\bibfnamefont {K.}~\bibnamefont
  {Uchino}},\ }\href@noop {} {\emph {\bibinfo {title} {Ferroelectric
  Devices}}}\ (\bibinfo  {publisher} {Marcel Decker, New York},\ \bibinfo
  {year} {2000})\BibitemShut {NoStop}%
\bibitem [{\citenamefont {Schlom}\ \emph {et~al.}(2007)\citenamefont {Schlom},
  \citenamefont {Chen}, \citenamefont {Eom}, \citenamefont {Rabe},
  \citenamefont {Streiffer},\ and\ \citenamefont {Triscone}}]{Schlom_2007}%
  \BibitemOpen
  \bibfield  {author} {\bibinfo {author} {\bibfnamefont {D.~G.}\ \bibnamefont
  {Schlom}}, \bibinfo {author} {\bibfnamefont {L.-Q.}\ \bibnamefont {Chen}},
  \bibinfo {author} {\bibfnamefont {C.-B.}\ \bibnamefont {Eom}}, \bibinfo
  {author} {\bibfnamefont {K.~M.}\ \bibnamefont {Rabe}}, \bibinfo {author}
  {\bibfnamefont {S.~K.}\ \bibnamefont {Streiffer}}, \ and\ \bibinfo {author}
  {\bibfnamefont {J.-M.}\ \bibnamefont {Triscone}},\ }\href@noop {} {\bibfield
  {journal} {\bibinfo  {journal} {Annual Review of Materials Research}\
  }\textbf {\bibinfo {volume} {{37}}},\ \bibinfo {pages} {589} (\bibinfo {year}
  {{2007}})}\BibitemShut {NoStop}%
\bibitem [{\citenamefont {Noeth}\ \emph {et~al.}(2008)\citenamefont {Noeth},
  \citenamefont {Yamada}, \citenamefont {Tagantsev},\ and\ \citenamefont
  {Setter}}]{Noeth_jap_2008}%
  \BibitemOpen
  \bibfield  {author} {\bibinfo {author} {\bibfnamefont {A.}~\bibnamefont
  {Noeth}}, \bibinfo {author} {\bibfnamefont {T.}~\bibnamefont {Yamada}},
  \bibinfo {author} {\bibfnamefont {A.~K.}\ \bibnamefont {Tagantsev}}, \ and\
  \bibinfo {author} {\bibfnamefont {N.}~\bibnamefont {Setter}},\ }\href
  {\doibase 10.1063/1.2999642} {\bibfield  {journal} {\bibinfo  {journal}
  {Journal of Applied Physics}\ }\textbf {\bibinfo {volume} {104}},\ \bibinfo
  {eid} {094102} (\bibinfo {year} {2008})}\BibitemShut {NoStop}%
\bibitem [{\citenamefont {Berge}\ \emph {et~al.}(2008)\citenamefont {Berge},
  \citenamefont {Norling}, \citenamefont {Vorobiev},\ and\ \citenamefont
  {Gevorgian}}]{Berge_jap_2008}%
  \BibitemOpen
  \bibfield  {author} {\bibinfo {author} {\bibfnamefont {J.}~\bibnamefont
  {Berge}}, \bibinfo {author} {\bibfnamefont {M.}~\bibnamefont {Norling}},
  \bibinfo {author} {\bibfnamefont {A.}~\bibnamefont {Vorobiev}}, \ and\
  \bibinfo {author} {\bibfnamefont {S.}~\bibnamefont {Gevorgian}},\ }\href
  {\doibase 10.1063/1.2896585} {\bibfield  {journal} {\bibinfo  {journal}
  {Journal of Applied Physics}\ }\textbf {\bibinfo {volume} {103}},\ \bibinfo
  {eid} {064508} (\bibinfo {year} {2008})}\BibitemShut {NoStop}%
\bibitem [{\citenamefont {Fukada}(2000)}]{Fukada}%
  \BibitemOpen
  \bibfield  {author} {\bibinfo {author} {\bibfnamefont {E.}~\bibnamefont
  {Fukada}},\ }\href@noop {} {\bibfield  {journal} {\bibinfo  {journal} {{IEEE
  Transactions on Ultrasonics, Ferroelectrics, and Frequency Control}}\
  }\textbf {\bibinfo {volume} {{47}}},\ \bibinfo {pages} {1277 } (\bibinfo
  {year} {{2000}})}\BibitemShut {NoStop}%
\bibitem [{\citenamefont {Kochervinskii}(2003)}]{Kochervinskii}%
  \BibitemOpen
  \bibfield  {author} {\bibinfo {author} {\bibfnamefont {V.}~\bibnamefont
  {Kochervinskii}},\ }\href@noop {} {\bibfield  {journal} {\bibinfo  {journal}
  {{Crystallography Reports}}\ }\textbf {\bibinfo {volume} {{48}}},\ \bibinfo
  {pages} {649 } (\bibinfo {year} {{2003}})}\BibitemShut {NoStop}%
\bibitem [{\citenamefont {Landolt}\ and\ \citenamefont
  {Bornstein}(1993)}]{Landolt-Bornstein}%
  \BibitemOpen
  \bibfield  {author} {\bibinfo {author} {\bibfnamefont {H.~H.}\ \bibnamefont
  {Landolt}}\ and\ \bibinfo {author} {\bibfnamefont {R.}~\bibnamefont
  {Bornstein}},\ }\href@noop {} {\emph {\bibinfo {title} {Numerical Data and
  Functional Relationships in Science and Technology}}}\ (\bibinfo  {publisher}
  {Springer},\ \bibinfo {year} {1993})\BibitemShut {NoStop}%
\bibitem [{\citenamefont {Yamada}(1972)}]{Yamada_BTO}%
  \BibitemOpen
  \bibfield  {author} {\bibinfo {author} {\bibfnamefont {T.}~\bibnamefont
  {Yamada}},\ }\href@noop {} {\bibfield  {journal} {\bibinfo  {journal}
  {Journal of Applied Physics}\ }\textbf {\bibinfo {volume} {43}},\ \bibinfo
  {pages} {328} (\bibinfo {year} {1972})}\BibitemShut {NoStop}%
\bibitem [{\citenamefont {Tagantsev}, \citenamefont {Courtens},\ and\
  \citenamefont {Arzel}(2001)}]{Tagantsev_STO}%
  \BibitemOpen
  \bibfield  {author} {\bibinfo {author} {\bibfnamefont {A.~K.}\ \bibnamefont
  {Tagantsev}}, \bibinfo {author} {\bibfnamefont {E.}~\bibnamefont {Courtens}},
  \ and\ \bibinfo {author} {\bibfnamefont {L.}~\bibnamefont {Arzel}},\
  }\href@noop {} {\bibfield  {journal} {\bibinfo  {journal} {Phys. Rev. B}\
  }\textbf {\bibinfo {volume} {64}},\ \bibinfo {pages} {224107} (\bibinfo
  {year} {2001})}\BibitemShut {NoStop}%
\bibitem [{\citenamefont {Pertsev}, \citenamefont {Zembilgotov},\ and\
  \citenamefont {Tagantsev}(1999)}]{Pertsev_Ferroelectrics_1999}%
  \BibitemOpen
  \bibfield  {author} {\bibinfo {author} {\bibfnamefont {N.~A.}\ \bibnamefont
  {Pertsev}}, \bibinfo {author} {\bibfnamefont {Z.~G.}\ \bibnamefont
  {Zembilgotov}}, \ and\ \bibinfo {author} {\bibfnamefont {A.~K.}\ \bibnamefont
  {Tagantsev}},\ }\href@noop {} {\bibfield  {journal} {\bibinfo  {journal}
  {Ferroelectrics}\ }\textbf {\bibinfo {volume} {223}},\ \bibinfo {pages} {79}
  (\bibinfo {year} {1999})}\BibitemShut {NoStop}%
\end{thebibliography}%
\end{document}